\documentstyle{elsart}
\begin{document}
\begin{frontmatter}
\title{Optical von Neumann measurement}
\author{G. M. D'Ariano and M. F. Sacchi}
\address{Dipartimento di Fisica \lq\lq A. Volta\rq\rq, via Bassi 6, 
I-27100 Pavia, ITALY\\
{\tt DARIANO@PV.INFN.IT} \\
{\tt MSACCHI@PV.INFN.IT} \\ 
{\bf 1993 PACS number(s):} 42.50.-p, 05.30.-d, 02.90.+p}
\begin{abstract}
We present an optical scheme that realizes the standard von 
Neumann measurement model, providing an indirect measurement 
of a quadrature of the field with controllable 
Gaussian state-reduction. The scheme is made of simple optical 
elements, as laser sources, beam splitters, and phase sensitive 
amplifiers, along with a feedback mechanism that uses a Pockels cell. 
We show that the von Neumann measurement is achieved without the 
need of working in a ultra-short pulsed regime.
\end{abstract}
\end{frontmatter}
In the last chapter of his book\cite{vn55} von Neumann 
formulated a measurement scheme for the position $\hat q$ of a 
particle based on a coupling with another particle. 
The interaction Hamiltonian between the two particles---object 
and probe---is of the form $\hat H_I=\hat q\hat P$, product of the 
object's position $\hat q$ with the probe's momentum $\hat P$. 
It is switched on with a very strong coupling and for a very short 
time, and immediately afterwards a measurement of the probe-particle 
position $\hat Q$ is performed. By shifting the probe's position 
$\hat Q$ by anamount proportional to the object's position 
$\hat q$, the coupling correlates the object's position with 
the probe's ``pointer observable'' $\hat Q$, 
through which the object's position is obtained, thus leaving 
the particle available for a forthcoming measure. 
\par Originally, von Neumann introduced his model in order to discuss 
the repeatability hypothesis suggested by the Compton-Simons 
experiment. After, it remained as a reference point for theoretical 
models of repeatable measurements, an ideal ``gedanken 
microscope'' with controllable disturbance on the system 
(see, for example, Ref. \cite{Ozawa}, where the von 
Neumann model is considered in relation to the problem of position 
measurements below the standard quantum limit). 
\par Is it possible to achieve this model experimentally? As a 
particle Hamiltonian, the $\hat q\hat P$ interaction is rather 
artificial. However, we will show that in the domain of quantum 
optics, one can achieve the von Neumann measurement (i. e. with 
the same probability distribution and the same ``state-reduction'') 
without the need of either realizing the precise form of the 
von Neumann Hamiltonian, or of experimentally achieving the impulsive 
limit: and from this prototype scheme for the von Neumann 
measurement we will have learned an interesting lesson useful 
for future quantum measurement engineering. 
The scheme we present is made of simple optical 
elements, as laser sources, beam splitters, and phase sensitive 
amplifiers, along with a feedback mechanism that uses a Pockels cell. 
We will employ a ``pre-amplification'' of the signal state 
and a ``pre-squeezing'' of the probe state: this is the basic idea 
to improve the quality of a quantum measurement \cite{jhs}, that has 
already been implemented in the realm of back-action evading measurements 
\cite{lap,ou,per,benc}.
\par For a single mode of the radiation field, the optical observables 
that correspond to particle position $\hat q$ and momentum $\hat p$ 
are represented by any two conjugated quadratures $\hat x_{\phi}$, and 
$\hat x_{\phi +\pi /2}$, with commutator $[\hat x_{\phi},\hat x_
{\phi +\pi /2}]=i/2$, 
the generic quadrature being defined as follows
\begin{eqnarray}
\hat x_{\phi}={{1}\over{2}}\left(a^{\dag}e^{i\phi}+ae^{-i\phi}\right)\;.
\label{qua}
\end{eqnarray}
The quadrature $\hat x_{\phi}$ can be ideally measured by means of a 
homodyne detector, in the limit of strong coherent local oscillator 
(LO), $\phi$ being the phase of the signal mode relative to the LO
\cite{darank}.
In Eq. (\ref{qua}), $a$ and $a^{\dag}$ are the 
bosonic annihilation and creation operators of the field mode of 
interest, with commutation $[a,a^{\dag}]=1$. 
\par Now we need to settle the general theoretical framework for describing 
repeatable measurements. In order to have a measurement that does not 
completely destroy the state that the system had before the 
measurement, the scheme must involve a probe 
that interacts with the system and later is ``measured'' to yield 
information on the original state of the system \cite{Ozawa}. 
This {\em indirect} measurement scheme is completely specified once the 
following ingredients are given: i) the unitary operator 
$\hat U$ that describes the system-probe interaction; ii) the state 
$|\varphi\rangle$ of the probe before the interaction; iii) the 
observable $\hat X$ which is measured on the probe. 
If, at the end of the system-probe interaction, one now considers 
another measurement on the system---say the ideal measurement of an 
observable $\hat Y$ (both $\hat X$ and $\hat Y$ 
have continuous spectrum, with eigenvectors $|x\rangle$ and 
$|y\rangle$, respectively), then the conditional probability 
density $p(y|x)$---of getting a result $y$ from the second 
measurement given the result of the first one being $x$---can 
again be written in terms of the Born's rule 
$p(y|x)dy=\langle y|\hat\varrho_x|y\rangle$ upon 
defining a ``reduced state'' $\hat\varrho_x$ as follows
\begin{eqnarray}
\hat\varrho_x=\frac{\hat\Omega(x)\hat\varrho\hat\Omega^{\dag}(x)}{
\mbox{Tr}[\hat\varrho\hat\Omega^{\dag}(x)\hat\Omega(x)]}
\;,\label{instr}
\end{eqnarray}
where the system operator $\hat\Omega(x)$ is given by 
\begin{eqnarray}
\hat\Omega(x) =\langle x|\hat U|\varphi\rangle\;,\label{Om}
\end{eqnarray}
and 
\begin{eqnarray}
d\hat\mu(x)=\hat\Omega^{\dag}(x)\hat\Omega(x)dx\;
\label{PomOm}
\end{eqnarray}
is the ``probability operator-valued measure'' (POM) of the apparatus 
\cite{Ozawa}, which provides the Born's rule for the measurement as 
follows \begin{eqnarray}
p(x)dx=\mbox{Tr}[\hat\varrho d\hat\mu(x)]
\;.\label{POMBorn}
\end{eqnarray}
Eqs. (\ref{instr}-\ref{POMBorn}) 
are the most general form of the state-reduction and of 
the Born's rule for a ``pure'' or ``quasi-complete'' measurement, 
namely a measurement that leaves pure states as pure 
(due to the pure state preparation of the probe). Apart from an 
irrelevant phase factor, the non-unitary {\em reduction operator}
$\hat\Omega(x)$ 
uniquely characterizes the quantum measurement, and two 
measurements that have the same operator $\hat\Omega(x)$ will be 
considered as identical, both having the same probability density 
(\ref{POMBorn}) and the same state-reduction (\ref{instr}). On the 
other hand, the fact that many measurements can share the same 
POM $d\hat\mu(x)$---while having different state reduction---is 
immediately apparent from the fact that a unitary transformation 
of the reduction operator $\hat\Omega(x)\to\hat\Omega '(x)
=\hat V(x)\hat\Omega (x)$ changes only the state-reduction, 
but leaves the POM (\ref{PomOm}) invariant. A unitary transformation 
$\hat V(x)$ that depends on the result $x$ of a measurement is the 
quantum mechanical description of a feedback mechanism, which in turn 
represents the easiest way of engineering a prescribed (admissible) 
state reduction.\par In the above framework, the standard von 
Neumann measurement model is given by the Gaussian reduction operator
\begin{eqnarray}
\hat\Omega (q)= \left( {1\over 2\pi\Delta^2}\right)^{1/4}
\exp\left[ -{(q-\hat q)^2\over 4\Delta ^2}\right]\;,
\label{vnom}
\end{eqnarray}
which results from an impulsive interaction Hamiltonian $\hat q\hat P$ 
with the probe-particle prepared in a Gaussian wave packet. From Eqs. 
(\ref{PomOm}) and (\ref{POMBorn}) it follows that the experimental 
probability density $p(q)=\mbox{Tr}[\hat\varrho\hat\Omega^{\dag}(q)
\hat\Omega(q)]$ is just a Gaussian convolution of the ideal 
probability distribution 
$\langle q|\hat\varrho|q\rangle$, with additional r.m.s. noise 
given by $\Delta$ (for interaction time $\tau=\kappa^{-1}$ 
with $\kappa$ the interaction strength, $\Delta^2$ is simply given by 
the variance of the probe-particle Gaussian wave-packet).
\par Now we present the quantum-optical scheme 
that performs the standard von Neumann measurement of the quadrature 
$\hat x_{\phi}$ of the radiation field in Eq. (\ref{qua}). 
In the following we will consider a fixed phase $\phi$, using 
the short notation $\hat x=\hat x_{\phi}$, $\hat y=\hat x_{\phi +\pi/2}$. 
In a fully optical measurement scheme the simplest choice for a 
measuring probe is just another mode of the field. We consistently use 
capital letters for the probe operators: thus, $\hat A$ and $\hat
A^{\dag}$ will denote the annihilation and creation operators 
of the probe mode, whereas $\hat X$ and $\hat Y$ will be used 
to represent any couple of conjugated quadratures of the probe 
for fixed phase $\phi'$. With this notation, the optical equivalent 
of the standard von Neumann Hamiltonian (for indirectly measuring 
the quadrature $\hat x$ by probing $\hat X$) is given by
\begin{eqnarray}
\hat H=\hat x\hat Y\;.\label{vn}
\end{eqnarray}
Notice that the choice of the phases $\phi$ and $\phi'$ is totally
free, and is ultimately related to the definition itself of the
annihilation and creation operators of the two modes.
From the definition (\ref{qua}) of $\hat x_{\phi}$ we can immediately 
see that, independently on the frequency of the two field modes, the 
Hamiltonian (\ref{vn}) cannot be realized in the rotating wave 
approximation, due to the counter-rotating terms $\hat A\hat a$ and
$\hat A^{\dag}\hat a^{\dag}$. On the other hand, an impulsive
realization of this Hamiltonian, as in the original formulation of von
Neumann, again is not feasible in the optical domain, because it
would require switching the interaction faster than the optical 
frequency. However, as we will show in the following, we don't 
need to realize the Hamiltonian (\ref{vn}) in order to achieve 
the von Neumann measurement. 
\par Instead of the Hamiltonian (\ref{vn}) we consider the interaction
of the two field modes at a beam splitter. This is described by 
the unitary evolution operator \cite{darank}
\begin{eqnarray}
\hat U=\exp{\left[ \mbox{atan} \sqrt{{1-\eta\over \eta}}\left( ab^{\dag}
-a^{\dag}b\right) \right] }\;.\label{UBS}
\end{eqnarray}
The unitary evolution operator (\ref{UBS}) has no counter-rotating 
terms: in the following we will take both modes at the same 
frequency, so that the operator (\ref{UBS}) will retain its 
time-independent form also in 
the interaction picture (the simple form of the operator 
(\ref{UBS}) holds for an appropriate choice of the modal phases, 
which can be achieved by just changing optical path lengths). 
Expressed as a function of the field quadratures, the unitary operator 
$\hat U$ reads
\begin{eqnarray}
\hat U=\exp{\left[2i \mbox{atan} \sqrt{{1-\eta\over \eta}}\left( 
\hat y\hat X -\hat x\hat Y\right) \right] }\;.\label{UXY}
\end{eqnarray}
The operator in Eq. (\ref{UXY}) can be conveniently factorized into 
the product of elemental unitary evolutions by exploiting the 
realization of the $su(2)$ algebra
$\hat J_+\equiv 2i\hat y \hat X$,
$\hat J_-\equiv 2i\hat x\hat Y$,
$\hat J_z\equiv i(\hat X\hat Y-\hat x\hat y)$, where one can easily
verify the $su(2)$ commutation relations
$[\hat J_+,\hat J_-]=2\hat J_z$, $[\hat J_z,\hat
J_{\pm}]=\pm \hat J_{\pm}$.
Using the BCH formula for the $SU(2)$ group \cite{notaBCH},
the operator $\hat U$ can be written as follows
\begin{eqnarray}
\hat U=e^{2i \sqrt{{1-\eta\over \eta}} \hat y\hat X }
\eta^{i\left(\hat x\hat y-\hat X\hat Y \right) }
e^{-2i \sqrt{{1-\eta\over \eta}} \hat x\hat Y}
\;.\label{fac}
\end{eqnarray}
The last factor on the right of Eq. (\ref{fac}) has the same form of
the von Neumann unitary evolution for Hamiltonian (\ref{vn}).
The physical meaning of other two factors will become clear after 
evaluating the reduction operator $\hat \Omega (x)$ corresponding to
the unitary evolution 
in Eq. (\ref{fac}).
\par Let $|\varphi\rangle$ be the state preparation of
the probe mode before the measurement, (the state of the field mode 
$A$ that enters one port of the beam splitter), and let us denote by
$|x\rangle$ the eigenvectors of the quadrature $\hat
X$ effectively measured at one output port of the beam splitter 
by means of a homodyne detector. Then, $\hat\Omega(x)$ 
can be evaluated through the following steps
\begin{eqnarray}
&&\hat \Omega (x)=
\langle x|e^{2i\sqrt{{1-\eta\over \eta}}\hat X\hat y}\eta^{i(\hat x \hat
y-\hat X\hat Y)}
e^{-2i\sqrt{{1-\eta\over \eta}}\hat x\hat Y}|\varphi\rangle \nonumber \\
&&= e^{2i\sqrt{{1-\eta\over \eta}}x\hat y}\eta^{i\hat x \hat
y}\langle x|
\eta^{-i\hat X\hat Y}
e^{-2i\sqrt{{1-\eta\over \eta}}\hat x\hat Y}|\varphi\rangle  \nonumber\\
&&= e^{2i\sqrt{{1-\eta\over \eta}}x\hat y}\eta^{i\hat x \hat
y}e^{-\ln \eta ^{1/2}x\partial_{x}}e^{-\sqrt{{1-\eta\over
\eta}}\hat x\partial_x}\varphi(x) \nonumber\\&&=
\hat D_{a}^{\dag}\left(\sqrt{{1-\eta\over \eta}}x\right)\,
\hat S_{a}^{\dag}\left(\ln \eta^{1/2}\right)\ \times \nonumber\\ 
&&\times\ \eta^{-1/4}\,\varphi\left[\eta^{-1/2} 
\left(x-(1-\eta)^{1/2}\hat x \right) \right] 
\;,\label{om1}   
\end{eqnarray}
where $\hat S_{a}(r)$ and $\hat D_{a}(\alpha)$ denote the squeezing and
displacement operators of the mode $a$, namely
\begin{eqnarray}
\hat S_{a}(r)=e^{-ir\left(\hat x\hat y+\hat y\hat x \right) }\;,
\qquad \hat D_{a}(\alpha)=e^{\alpha a^{\dag}-\bar\alpha a}\;,
\end{eqnarray}
and we used the quadrature differential representation
\begin{eqnarray}
\langle x|f(\hat X,\hat Y) |\varphi\rangle= 
f\left(x,-{i\over 2}\partial_{x}\right)\varphi(x) \,;\quad
\varphi(x)\equiv\langle x|\varphi\rangle\;.
\end{eqnarray}
The squeezing and displacement unitary operators that appear in 
the last step of Eq. (\ref{om1}) represent an additional back action 
from the measurement, i. e. they just change the state-reduction by an 
additional unitary evolution, but they do not change the POM, 
which for the reduction operator (\ref{om1}) is given by
\begin{eqnarray}
d\hat\mu_{\eta}(x)&=& dx\,\hat \Omega^{\dag} (x)\hat \Omega (x)
\nonumber \\&=&
dx\,\eta^{-1/2}\left |\varphi\left[\eta^{-1/2} \left(x-(1-\eta)^{1/2}\hat x 
\right) \right]\right |^2\;.\label{pom1}
\end{eqnarray}
For very high reflectivity at the beam splitter
$\eta\rightarrow 0$ and with the probe prepared in the vacuum
state $|\varphi\rangle\equiv|0\rangle$, Eq. (\ref{pom1}) would approach
the Gaussian von Neumann POM from Eqs. (\ref{PomOm}) and (\ref{vnom}) with 
variance $\Delta=\sqrt{\eta}/2$. However, the reduction operator 
(\ref{om1}) is still different from that in Eq. (\ref{vnom}), and in 
order to make them equal we need to remove the squeezing and the 
displacement back-action terms. 
The displacement term is a unitary transformation that depends on the 
measurement outcome, and hence it can be compensated by an appropriate 
feedback device. On the other hand, the 
squeezing term can be balanced by an inverse squeezing transformation 
of the mode $a$ performed after the displacing feedback: this will be 
the last transformation on the mode $a$, and we will refer to it as 
{\em back-squeezing}. For vanishing $\eta$ one would need 
increasingly large back-squeezing, and it may be more convenient to 
compensate the vanishing $\eta$ by 
squeezing the probe state $|\varphi\rangle$. In fact, squeezing 
transforms the quadrature $\hat x$ as follows
\begin{eqnarray}
\hat x \rightarrow\hat S^{\dag}_{a}(r)\,\hat x \,\hat S_{a}(r)=e^r \hat x\;.
\end{eqnarray}
Hence the factor $(1-\eta)^{1/2}$ in the POM (\ref{pom1}) can be removed by
{\em pre-squeezing} the initial state of the system with squeezing 
parameter $r=-{1\over2}\ln(1-\eta)$. 
Such pre-squeezing modifies the reduction operator $\hat\Omega(x)$ 
in Eq. (\ref{om1}) into the following one
\begin{eqnarray}
\hat\Omega(x)\to\hat\Omega(x)=
\,\eta^{-1/4}\varphi\left[\eta^{-1/2} \left(x-\hat x 
\right) \right]\;,\label{om2}
\end{eqnarray}
where now we have changed the back-squeezing as follows 
\begin{eqnarray}
\hat S_{a}^{\dag}(\ln \eta^{1/2})& \longrightarrow &
\hat S_{a}^{\dag}(\ln \eta^{1/2}) \hat S_{a}[\ln(1-\eta)^{-1/2}] \nonumber \\
&=&\hat S_{a}^{\dag}\left[{1\over 2}\ln (\eta(1-\eta))\right]  \;.
\end{eqnarray}
Then, in order to get a tunable variance for the reduction  operator, one
can change the state preparation $|\varphi\rangle$ of the probe. For
the squeezed vacuum
\begin{eqnarray}
|\varphi\rangle=\hat S_{A}(\ln\sigma^{1/2})|0\rangle\;,\label{vac}
\end{eqnarray}
the reduction operator (\ref{om2}) becomes
\begin{eqnarray}
\hat\Omega(x)=\left({2\over\pi\eta\sigma}\right)^{{1\over4}}
\exp \left[-{(x-\hat x)^2\over\eta\sigma}\right]\;,\label{om3}
\end{eqnarray}
and the operator $\hat\Omega(x)$ in Eq. (\ref{om3}) is of the same
form of the von Neumann one in Eq. (\ref{vnom}), with 
$\Delta=\sqrt{\eta\sigma}/2$. 
\par The experimental set-up to perform the optical von Neumann
measurement is sketched in Fig. 1.
\begin{figure}[htb]
\begin{center}
\vspace{60mm}
\end{center}
\caption{Outline of the proposed experimental setup to realize a von
Neumann measurement of a quadrature of the electromagnetic field. BS
denote a beam splitter; PC$(\theta)$ denotes a Pockels cell with
transmissivity $\theta$.}
\label{f:vonfig}
\end{figure}
\par\noindent 
The pre-squeezing and back-squeezing transformation described 
by the unitary operators
\begin{eqnarray}
\hat S^1_{a}\left[-{1\over 2}\ln (1-\eta)\right]\;,
\qquad
\hat S^2_{a}\left[{1\over 2}\ln (\eta(1-\eta))\right]\;,
\end{eqnarray}
are the two extremal steps of the sequence of optical operations on the
system mode.  They can be accomplished by two phase-sensitive
amplifiers (PSA) \cite{Yuen} with gains 
$G_{1}=(1-\eta)^{-1}$ and $G_{2}=\eta(1-\eta)$, respectively.
[The PSA ideally amplifies the quadratures of the field
with a phase-dependent gain, namely
$\hat{x}'_{\phi}=G^{-1/2}\hat x_{\phi}$, $\hat{x}'_{\phi +\pi/2}
=G^{1/2}\hat x_{\phi +\pi/2}$. 
It can be attained through degenerate three- or 
four-wave mixing.]
In the same way the probe state (\ref{vac}) can be achieved using a
third PSA that amplifies an input vacuum field with gain $G_{3}=\sigma$. 
After the first squeezing, the state of the $a$ mode is entangled with 
the squeezed vacuum state (\ref{vac}) of the $A$ mode 
through the beam splitter with transmissivity $\sqrt{\eta}$, and at 
the reflected output beam the quadrature 
$\hat X$ is homodyne detected. The displacement $\hat D\left(\sqrt{{
1-\eta\over \eta}}x\right)$ is achieved by combining the transmitted beam 
with a strong coherent LO
$|\beta\rangle \ (\beta \rightarrow \infty)$ in a beam splitter 
with a transmissivity $\theta \rightarrow 1$, such that 
$|\beta|\sqrt{1-\theta}=\sqrt{{1-\eta\over \eta}}x$ (the LO
is at the frequency of the signal mode $a$). The parametric dependence
on the homodyne outcome $x$ is carried out by driving the LO 
with the homodyne photocurrent, for example by stimulating
the laser that provides the LO by the photodetection current itself.
However, this method is expected to fail for small ``photocurrents''
$x$, because it would bring the LO laser below threshold, thus loosing
the phase of $\beta$. A better way to achieve this feedback is to provide 
a current-dependent transmissivity $\theta(x)$ for the beam-splitter,
making use, for example, of a Pockels cell, and working in the
linearity regime $\theta\propto x$ of the cell [a similar feedback mechanism 
has been experimentally implemented in Ref. \cite{Demartini}].
Of course, good phase coherence between the PSA pumps and the LO
may be technically difficult to achieve. Finally,
also notice that the quadrature phase $\phi$ can be changed in many 
different ways by tuning any one of the relative phase-shifts between 
the pumps and the LO.
\par In conclusion, we have presented a quantum-optical scheme
that realizes the standard von Neumann model, a model for repeatable
quantum measurements with controlled state-reduction. Our scheme uses 
simple optical elements, like beam-splitters and squeezers. 
We have seen that, contrarily to the customary modeling of 
repeatable measurements, there is no need of 
working in a ultra-short pulsed regime. We have also seen how the
precise form of the state reduction can be engineered by means 
of a feedback mechanism that uses a Pockels cell: we think that
this method can be of use in more general situations, for controlling
the back-action of a quantum measurement.
Finally we hope that our scheme will be implemented experimentally, 
and will be of help for a deeper understanding of the physical mechanisms
underlying quantum measurements.


\begin{thebibliography}{99}
\bibitem{notaBCH} Notice that the fact that our realization 
of the $su(2)$ algebra does 
not preserve hermitian conjugation is irrelevant for the group 
multiplication, as in the BCH formula we are actually using the 
complexification of the group.
\end{thebibliography}
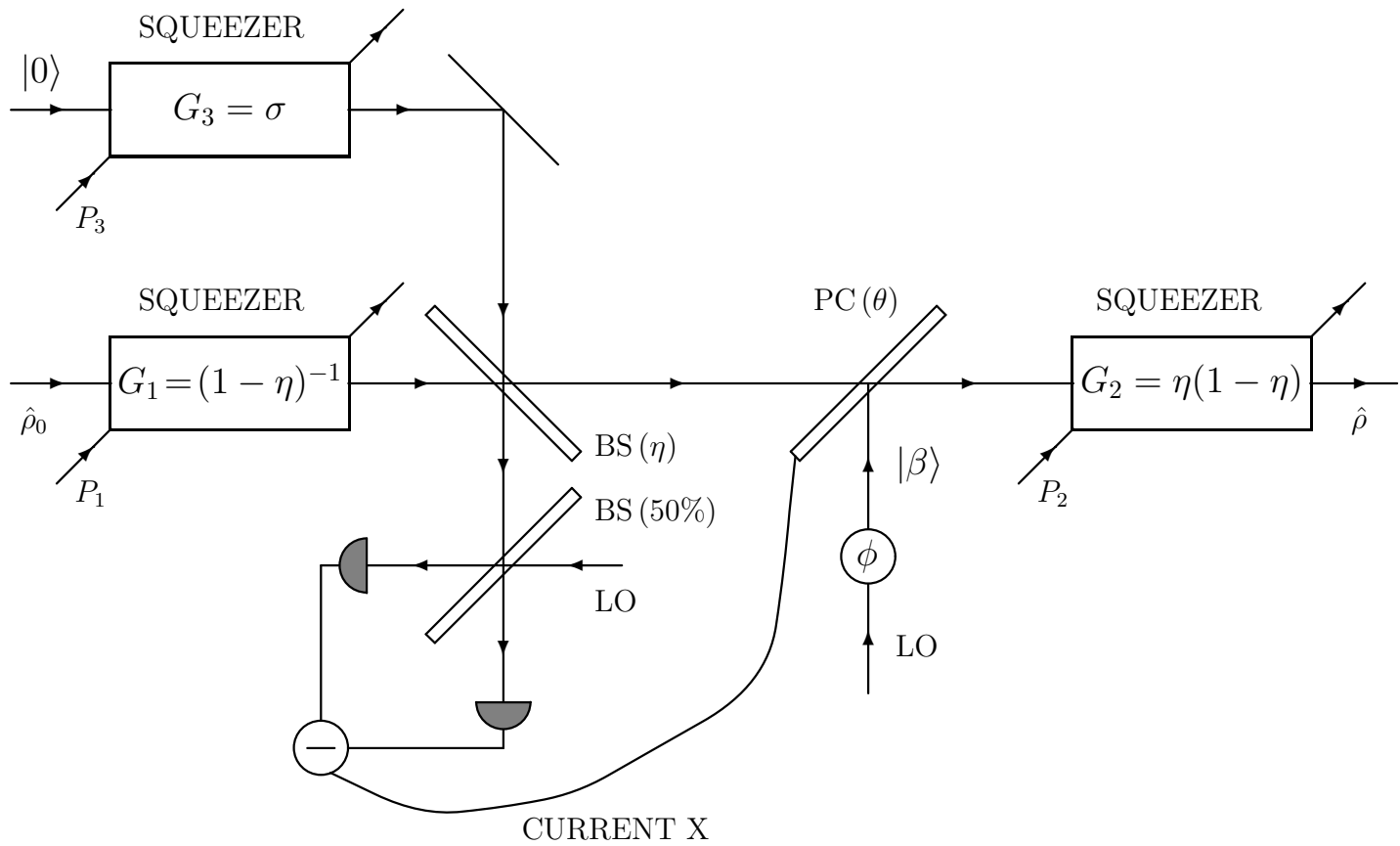
\end{document}